\documentclass[preprint,nofootinbib,showpacs]{revtex4-1}
\usepackage{amsmath,amsfonts,amssymb,latexsym}
\usepackage{amsmath}
\usepackage{amsfonts}
\usepackage{amssymb}
\usepackage{graphicx}

\def \be{\begin{equation}}
\def \ee{\end{equation}}
\def \bq{\begin{eqnarray}}
\def \eq{\end{eqnarray}}
\def \beq{\begin{eqnarray*}}
\def \eeq{\end{eqnarray*}}

\begin{document}

\title{{\Large Inflation Without a Trace of Lambda}}
\author{John D. Barrow}
\email{J.D.Barrow@damtp.cam.ac.uk}
\affiliation{DAMTP, Centre for Mathematical Sciences \\
Wilberforce Rd., Cambridge University \\
Cambridge CB3 0WA, UK}
\author{Spiros Cotsakis}
\email{skot@aegean.gr}
\affiliation{Department of Mathematics\\
College of Engineering and Technology \\
American University of the Middle East\\
Kuwait }
\begin{abstract}
\noindent We generalise Einstein's formulation of the traceless Einstein
equations to $f(R)$ gravity theories. In the case of the vacuum traceless
Einstein equations, we show that a non-constant Weyl tensor leads via a
conformal transformation to a dimensionally homogeneous (`no-scale') theory
in the conformal frame with a scalar field source that has an exponential
potential. We then formulate the traceless version of $f(R)$ gravity, and we
find that a conformal transformation leads to a no-scale theory  conformally
equivalent to general relativity and a scalar field $\phi $ with a potential
given by the scale-invariant form: $V(\phi )=\frac{D-2}{4D}Re^{-\phi }$,
where $\phi =[2/(D-2)]\ln f^{\prime }(R)$. In this theory, the cosmological
constant is a mere integration constant, statistically distributed in a
multiverse of independent causal domains, the vacuum energy is another
unrelated arbitrary constant, and the same is true of the height of the
inflationary plateau present in a huge variety of potentials. Unlike in the
conformal equivalent of full general relativity, flat potentials are found
to be possible in all spacetime dimensions for polynomial lagrangians of all
orders. Hence, we are led to a novel interpretation of the cosmological
constant vacuum energy problem and have accelerated inflationary expansion
in the very early universe with a very small cosmological constant at late
times for a wide range of no-scale theories. Fine-tunings required in
traceless general relativity or standard non-traceless $f(R)$ theories of
gravity are avoided. We show that the predictions of the scale-invariant
conformal potential are completely consistent with microwave background
observational data concerning the primordial tilt and the tensor-to-scalar
ratio.
\end{abstract}

\date{July, 2020}
\maketitle
\tableofcontents

\newpage

\section{Introduction}

The vacuum Einstein equations $R_{\mu \nu }-1/2g_{\mu \nu }R=0$ are scale
invariant in the sense that the Einstein tensor is dimensionally
homogeneous: each of the two terms on the left-hand side are homogeneous of
degree two in the derivatives of the metric and there is no length with
respect to which one can measure the size of any dimensionful constant. The
introduction of a cosmological constant into the energy density of the
(Lorentz-invariant) vacuum breaks scale invariance by adding an
energy-momentum tensor of the form $T_{\mu \nu }=-\Lambda g_{\mu \nu }$ to
the vacuum field equations \cite{lem} and permits the usual inflationary
cosmological solutions. However, it also creates a scale given by $\Lambda
=(8\pi G)^{-1/2}$, and hence a discrepancy between the observed energy
density of empty space, which is measured to be around $10^{-47}\text{GeV}%
^{4}$ using general relativity, and the expected vacuum field energy $%
\Lambda ^{4}/16\pi ^{2}\approx 2\times 10^{71}\text{GeV}^{4}$, from quantum
field theory. This is sometimes called the `cosmological constant problem'
\cite{BT, weinberg89, BShaw2010, BShaw}. By introducing $\Lambda $ into the
vacuum Einstein equations we introduce the possibility of inflationary
(accelerating) de Sitter solutions at the expense of an anomalously large
vacuum energy density.

The main proposal advanced in this paper is that this discrepancy points to
a fundamental failure of both general relativity and quantum field theory.
For they are unable to deal with the cosmological constant in a meaningful
way simply because $\Lambda $ is something extraneous to both theories. We
take it as a basic point of departure that the cosmological constant implies
that some new physics operates at a hyper-classical scale, not reducible to
either the classical domain of general relativity or that of quantum fields.
The ideas of the multiverse and the anthropic principle \cite{BT} contain
some of the spirit of the present work, but our starting point and
implementation are entirely different. It is one of the basic results of
this paper that the cosmological constant and the energy of the vacuum are
random and accidental attributes of our observable universe, indicative of
the existence of a much wider, unexplored structure operating on a truly
cosmological scale - that of containing the various \textit{presently}
causally disconnected regions. We show that although the cosmological
constant is an arbitrary one appearing in the field equations of the theory,
it can be equated to its presently observed value in our local, causally
connected region, provided there is a distribution of such values, one for
each such independent domain in the multiverse. Here we have in mind not one
region (like ours) containing local subregions that were causally
disconnected before the occurrence of inflation and became causally related
after inflation, but completely separate regions like those that are
constantly produced in the self-regenerating inflationary universe \cite%
{linde,barrow-cot,cot}. However, our model is not based or dependent on that
process.

In our approach, we can retain the presence of inflationary solutions in the
theory without automatically admitting the possibility of a vacuum energy
density because as we show in this paper both the cosmological constant and
the vacuum energy can be generally unrelated arbitrary constants. In Section
II, we describe formulation of the traceless version of the Einstein
equations, first proposed by Einstein, and we prove using the Weyl tensor
that the vacuum traceless Einstein equations include all the vacuum
solutions of general relativity plus all solutions having a non-constant
Weyl tensor (having a non-zero cosmological constant). As we show in Section
II.B for the first time, the vacuum traceless theory is conformally
equivalent to general relativity plus a self-interacting scalar field with
an exponential potential. This result uses in an essential way a new effect
that we highlight here, namely, the dimensional homogeneity of the traceless
equations, a property closely related (in a sense which we make precise) to
the traceless equations being `scale invariant'. We also discuss the
physical relevance of the exponential potential of the theory. In Section
III, which is the main Section of this work, after reviewing certain
relevant features of the standard $f(R)$ theory, we introduce and study a
new variant, \emph{the no-scale $f(R)$ gravity}. In Section III.B, we show
that the no-scale $f(R)$ theory bears a similar relation to standard $f(R)$
gravity as traceless GR does to standard GR with a cosmological constant, we
provide the integrability condition, making the cosmological constant an
arbitrary constant in this case. In Section III.C, we show that the no-scale
$f(R)$ theory in spacetime dimension $D$ is conformally equivalent to
general relativity with a scalar field $\phi $ with a potential given by a
scale-invariant form: $V(\phi )=\frac{D-2}{4D}Re^{-\phi }$, where $\phi
=[2/(D-2)]\ln f^{\prime }(R),$ and $f^{\prime }\equiv \partial f/\partial R$%
. This scalar field potential is quite different from the one we know from
the standard conformal relation of $f(R)$ gravity and general relativity
found in ref. \cite{ba-co88}. Its most interesting property is that, unlike
in the conformal equivalent of full general relativity, flat potentials in
no-scale theories are found to be possible in all spacetime dimensions for
polynomial lagrangians of all orders. In Section III.D, we discuss the
example of the no-scale $R+R^{n}$ theory, and we demonstrate how in this
theory the cosmological constant, the vacuum energy and the height of the
inflationary plateau are all unrelated arbitrary constants. Hence, no-scale
higher-order gravity, like traceless general relativity, avoids a
cosmological constant problem associated with an energy density of the
vacuum, and we are led to a statistical distribution of values of the
possible vacuum energies of individual components. In Section IV, we
calculate the slow-roll parameters and show that no-scale higher-order
gravity provides many possible inflationary models in agreement with
microwave background data, and avoids fine tuning problems. In particular,
we calculate the the primordial tilt and the tensor-to-scalar ratio for the
no-scale $R+R^{2}$ theory in four and general dimensions. The generic
prediction is that any future measurement of the form $r=12/(b^{2}N^{2}),b%
\neq 1$, will be in favor of the no-scale $R+R^{2}$ theory. We also
generalize these results to spacetime dimension $D$ and polynomial degree $n$
of the scalar field potential. We present our conclusions in Section V. In
the Appendix, we present the details of the conformal techniques leading to
the scalar field potentials for the no-scale general relativity (App. A1)
and no-scale $f(R)$ gravity (App. A2) respectively.

\section{No-scale general relativity}

We seek a situation where the cosmological constant problem is absent but
the possibility of inflation remains. We note that there is a way, first
suggested by Einstein \cite{einstein}, to cancel the effects of a
cosmological constant by using only the traceless parts of the Einstein
equations to define the theory of gravitation \cite{einstein,weinberg89,
ellis1, ellis2}. Before extending this approach to higher-order lagrangians,
we summarize the structure of traceless general relativity introduced by
Einstein and clearly expounded by Ellis in ref. \cite{ellis2}.

\subsection{General properties}

For any tensor $L_{\mu \nu }$, its traceless part in $D$ spacetime
dimensions is defined to be the new tensor,
\begin{equation}
\widehat{L}_{\mu \nu }=L_{\mu \nu }-\frac{1}{D}g_{\mu \nu }L,\quad L=\text{Tr%
}L_{\mu \nu },  \label{L}
\end{equation}%
so that $\widehat{L}=\text{Tr}\widehat{L}_{\mu \nu }=0$. For the Einstein
tensor, defined by $G_{\mu \nu }=R_{\mu \nu }-(1/2)g_{\mu \nu }R$, and
setting the Einstein constant $8\pi G/c^{4}=1$, we find that in $D$
dimensions its traceless part is \cite{einstein},
\begin{equation}
\widehat{G}_{\mu \nu }=R_{\mu \nu }-\frac{1}{D}g_{\mu \nu }R,  \label{tr0ein}
\end{equation}%
(since $1/D=1/2+(2-D)/2D$), so it coincides with the traceless Ricci tensor.
The traceless Einstein equations are then postulated to be
\begin{equation}
\widehat{G}_{\mu \nu }=\widehat{T}_{\mu \nu },\quad \widehat{T}_{\mu \nu
}=T_{\mu \nu }-\frac{1}{D}g_{\mu \nu }T,  \label{notrEE}
\end{equation}%
where $T_{\mu \nu }$ is the usual energy-momentum tensor of matter, and $T$
its trace. The vacuum traceless Einstein equations are scale invariant,
being dimensionally homogeneous, like the standard vacuum Einstein
equations. All the vacuum solutions of general relativity remain unchanged
and the nine traceless field equations are a subset of the full (ten)
Einstein equations. Crucially, however, although the stress tensor $T_{\mu
\nu }$ can have a trace, only the tracefree part gravitates\footnote{%
Thus, for example a perfect fluid with normalised 4-velocity $u_{\mu }$ has
stress tensor $\hat{T}_{\mu \nu }=(\rho +p)(u_{\mu }u_{\nu }+\frac{1}{4}%
g_{\mu \nu })$ with $\hat{T}=0.$}.

In the following, we shall mostly be interested in the vacuum equations, $%
\widehat{G}_{\mu\nu}=0$, which are obtained when $T_{\mu\nu}=0$. In the case
of a cosmological constant, $T_{\mu\nu}=-\Lambda g_{\mu\nu}$, the terms on
the right-hand side of the traceless equations (\ref{notrEE}) cancel each
other leading to the vacuum traceless equations as well. This means that in
the traceless theory, unlike in general relativity, the cosmological
constant has nothing to do with an energy density of the vacuum or a
constant term in the gravitational action. In fact, it can easily be shown
to be simply related to an arbitrary integration constant, $\lambda$, that
appears in the Einstein equations,
\begin{equation}
G_{\mu\nu}-\lambda g_{\mu\nu}=T_{\mu\nu},  \label{uniEE}
\end{equation}
with \cite{weinberg89, ellis1, ellis2},
\begin{equation}
R+T=-D\lambda.  \label{constraint}
\end{equation}
In this way, the traceless equations (\ref{notrEE}) reduce to the equivalent
system of equations (\ref{uniEE}), (\ref{constraint}).

To gain a better idea of the structure of the solution space of the vacuum
traceless Einstein equations, we can compare the Weyl tensor $%
C^{\rho\,}_{\,\,\,\alpha\beta\gamma}$ in this context to that of the vacuum
Einstein equations $R_{\mu \nu }=0$ (for simplicity we consider the case $D=4
$). Using the standard formula (equivalent to the Bianchi identities),
\begin{equation}
\nabla_\rho\, C^{\rho\,}_{\,\,\,\alpha\beta\gamma}=R_{\alpha[\gamma;\beta]}-%
\frac{1}{6}g_{\alpha[\gamma}R_{;\beta]},
\end{equation}
and the vacuum traceless equations $\widehat{G}_{\mu\nu}=0$, or equivalently
the system (\ref{uniEE}), (\ref{constraint}), we find that
\begin{equation}
\nabla_\rho\, C^{\rho\,}_{\,\,\,\alpha\beta\gamma}=0\quad \text{if and only
if}\quad \lambda=0.
\end{equation}
This result is important because it shows that unlike the vacuum Einstein
equations where the Weyl tensor is covariantly constant, the vacuum
traceless equations $\widehat{G}_{\mu\nu}=0$ have a more general solution
space which includes all vacuum solutions of general relativity ($\lambda=0$%
), but also many more solutions all of which have a non-constant Weyl
conformal curvature (corresponding to $\lambda\neq 0$.) Since the Weyl
tensor is a conformal invariant, we shall see in the next subsection why the
vacuum traceless Einstein theory is able to produce inflationary solutions
`without a trace of lambda'.

\subsection{The conformal potential and inflation}

Before we examine the inflationary structure of the vacuum traceless theory,
we make a comment about the inclusion of scalar fields in the traceless
general relativity context. Since for a scalar field $\phi$ with potential $%
V(\phi)$, the traceless part of its energy-momentum tensor, $\widehat{T}%
_{\mu\nu}=\partial_{\mu}\phi\partial_{\nu}\phi-(1/D)g_{\mu\nu}(\partial%
\phi)^{2}$, does not contain the potential $V(\phi)$ and we have $\widehat{T}%
=0$, it follows that in the traceless theory given by (\ref{notrEE}) the
scalar field $\phi$ gravitates but without a potential. Therefore, any
property of the solutions of the traceless field equations (\ref{notrEE})
which depends implicitly on the potential $V(\phi)$, such as the possibility
of slow-roll inflation, should probably be considered as an `off-shell'
effect requiring some sort of fine-tuning. 

We now move on to examine the conformal structure and possibility of
inflation in the vacuum traceless theory $\widehat{G}_{\mu\nu}=0$. We can
restrict the metric variations $\delta g_{\mu\nu}$ used in the action
principle by moving to a metric conformally related to $g_{\mu\nu}$ (that
enters in the vacuum traceless field equations), by introducing an analytic
function $f(R)$ by setting \cite{ba-co88}
\begin{equation}
\tilde{g}_{\mu\nu}=e^{\phi}g_{\mu\nu},\quad\phi=\frac{2}{D-2}\ln f^{\prime }.
\label{def}
\end{equation}
Then, following the steps in the Appendix (Part A), we find that the vacuum
traceless Einstein equations
\begin{equation}
\widehat{G}_{\mu\nu}=0,
\end{equation}
transform into,
\begin{equation}
\widetilde{\widehat{G}}_{\mu\nu}+\frac{D-2}{2}\nabla_{\mu}\nabla_{\nu}\phi=%
\tilde{T}_{\phi,\mu\nu}  \label{vac conf TEE1}
\end{equation}
where,
\begin{equation}
\tilde{T}_{\phi,\mu\nu}=\frac{D-2}{4}\left[ \nabla_{\mu}\phi\nabla_{\nu}\phi-%
\frac{1}{2}\tilde{g}_{\mu\nu}\left( \frac{2(D-1)}{D}(\nabla\phi
)^{2}+2V(\phi)\right) \right] ,  \label{matter tensor 2}
\end{equation}
with
\begin{equation}
V(\phi)=\frac{-\lambda}{(D-1)}e^{-\phi},  \label{potnew}
\end{equation}
where we have also used the constraint (\ref{constraint}) with $T=0$.

The equations (\ref{vac conf TEE1}), (\ref{matter tensor 2}) are scale
invariant in the sense of being dimensionally homogeneous: Each one of the
terms in the two sides of these equations has the same degree, namely two,
in the derivatives because they contain two derivatives of the metric or of
the scalar field (in the potential term we have substituted $R=-D\lambda $
from the constraint). Therefore, we can freely perform a scale
transformation $\phi \rightarrow a\phi +b$, with $a,b$ arbitrary constants
in the potential (and also redefine the location of the minimum of $V(\phi )$
by performing $V(\phi )\rightarrow V(\phi )+d$, where $d$ is any constant),
without spoiling the property of the equations being dimensionally
homogeneous. Hence, we end up with a conformal potential of the general form
$V(\phi )=ge^{-a\phi }$, with $g=-\lambda e^{b},a$ being arbitrary
constants. Note that when $\lambda =0$, the potential vanishes and we are
back to the case of vacuum general relativity.

Unfortunately, this generic prediction of the vacuum traceless general
relativity theory for inflation with an exponential potential is ruled out
both due to the fact that in this model there is no graceful exit from
inflation, but also because of its disagreement with current observations
(cf. eg., \cite{wein2}, Section 10.2).

\section{No-scale $f(R)$ gravity}

\subsection{Scale invariance in $f(R)$ gravity}

Adding a cosmological constant term is not the only way to break the scale
invariance of the vacuum Einstein equations. The addition of higher-order
curvature terms to the lagrangian also introduces a scale (see for example,
\cite{wein}, p. 153), and can lead to inflation in a natural way \cite{NT,
star, BO, JDB}. We consider the theory derived from a gravitational
lagrangian that is an arbitrary analytic function of the scalar curvature, $%
f(R),$ \cite{BO}. The field equations in vacuum, with $f^{\prime }=df/dR$,
derived by varying the action formed from the lagrangian $f(R),$ are
\begin{equation}
M_{\mu \nu }=f^{\prime }R_{\mu \nu }-\frac{1}{2}g_{\mu \nu }f-\nabla _{\mu
}\nabla _{\nu }f^{\prime }+g_{\mu \nu }\Box f^{\prime }=0.  \label{fofr1}
\end{equation}%
These equations are fourth order in the spacetime derivatives of the metric,
$g_{\mu \nu }$. The most interesting property they possess for our present
purposes is that they miss being scale invariant, in the sense of being
dimensionally homogeneous in the spacetime derivatives, simply because of
the presence of the term proportional to $g_{\mu \nu }f$. If this term were
absent, Eq. (\ref{fofr1}) would be uniform in scale in the same way that the
vacuum Einstein equations are for the spacetime metric $g_{\mu \nu }$ -- but
now containing second-order terms in the spacetime derivatives of the metric
as well as those of the `field' $f^{\prime }$. We further note that the
scale invariance breaking term containing $f$ is also present in the trace $%
M=\text{Tr}M_{\mu \nu }$, where for instance when $D=4$, we have
\begin{equation}
M=f^{\prime }R-2f+3\Box f^{\prime }.  \label{trace}
\end{equation}%
There is also the more suggestive form for these vacuum equations (assuming $%
f^{\prime }\neq 0$ for all $R$), first introduced in Ref. \cite{ba-co88},
for $M_{\mu \nu }$ which picks out the Einstein tensor explicitly:
\begin{equation}
M_{\mu \nu }=f^{\prime }G_{\mu \nu }+\frac{1}{2}g_{\mu \nu }(Rf^{\prime
}-f)-\nabla _{\mu }\nabla _{\nu }f^{\prime }+g_{\mu \nu }\Box f^{\prime }=0,
\label{fofr}
\end{equation}%
and also introduces the term proportional to $Rf^{\prime }-f$ which can
provide an inflationary potential. This potential, which was found in Ref.
\cite{ba-co88} to be proportional precisely to the scale invariance breaking
term $(Rf^{\prime }-f)f^{\prime -2}$, is important in the conformal picture
of general relativity with a self-interacting scalar field as source \cite%
{whitt}, as it allows an inflationary phase in cosmological solutions of
these theories \cite{ba-co88, maeda}.

\subsection{No-scale $f(R)$ gravity}

The question then arises as to whether the traceless version of Eq. (\ref%
{fofr1}) can solve the scale invariance breaking problem of the standard $%
f(R)$ gravity while maintaining the inflationary character of solutions of
the theory. In such a theory, like in the traceless version of general
relativity studied in the previous Section, there will be no particular
scale with respect to which one can measure the sizes of dimensionful
constants, it will be a \textit{no-scale} version of $f(R)$-theory.

Setting $\widehat{M}_{\mu\nu}=M_{\mu\nu}-(1/D)g_{\mu\nu}M,$ with $M=\text{Tr}%
M_{\mu\nu}$, we find
\begin{equation}
\widehat{M}_{\mu\nu}=f^{\prime}\left( \widehat{G}_{\mu\nu}-(f^{\prime})^{-1}%
\left( \nabla_{\mu}\nabla_{\nu}f^{\prime}-\frac{1}{D}g_{\mu\nu}\Box
f^{\prime}\right) \right) .  \label{Dzerotrace1}
\end{equation}
In contrast to the original higher-order field equations for $f(R)$ gravity (%
\ref{fofr1}), we see that there is now no term proportional to $g_{\mu\nu}f$
because it is cancelled by a similar term in the trace $M$. So in terms of
the notional scalar `field' $\Phi=(g_{\mu\nu},f^{\prime})$, the tensor $%
\widehat{M}_{\mu\nu}$ is scale invariant in the sense, like before, that it
is a homogeneous polynomial of degree two in the spacetime derivatives of
the field $\Phi$.

Following the traceless general relativity development, we postulate the
full traceless (or `no-scale') $f(R)$ field equations to be
\begin{equation}
\widehat{M}_{\mu\nu}=\widehat{T}_{\mu\nu},\quad\widehat{T}%
_{\mu\nu}=T_{\mu\nu }-\frac{1}{D}g_{\mu\nu}T,  \label{Dzerotrace2}
\end{equation}
where the left-hand side is given by (\ref{Dzerotrace1}). The stress tensor
of matter and radiation $T_{\mu\nu}$ on the right-hand side of this equation
will in general have a nonzero trace but only its traceless part will
gravitate. We note that by setting $T_{\mu\nu}=0,\,\,\text{or}\,\,-\Lambda
g_{\mu\nu}$, Eq. (\ref{Dzerotrace2}) becomes the vacuum, scale invariant
equation
\begin{equation}
\widehat{M}_{\mu\nu}=0,  \label{Dzerotrace2VAC}
\end{equation}
We will show that Eq. (\ref{Dzerotrace2}) is related to the standard $f(R)$%
-matter equations through a new, nontrivial integrability condition and, as
in traceless general relativity, eq. (\ref{constraint}), the cosmological
constant emerges an arbitrary integration constant. In this respect, the
traceless $f(R)$ situation is completely analogous to that of traceless
general relativity \cite{weinberg89,ellis2}. From (\ref{Dzerotrace2}), we
find by covariant differentiation that
\begin{equation}
\nabla^{\mu}\widehat{M}_{\mu\nu}=-\frac{1}{D}\nabla_{\nu}T,
\end{equation}
since we assume the usual conservation law, $\nabla^{\mu}T_{\mu\nu}=0$.
However, since (\ref{fofr1}) gives
\begin{equation}
\nabla^{\mu}{M}_{\mu\nu}=0,
\end{equation}
it follows from the definition of the traceless tensor $\widehat{M}_{\mu\nu}$
that
\begin{equation}
\nabla^{\mu}\widehat{M}_{\mu\nu}=-\frac{1}{D}\nabla_{\nu}M,
\end{equation}
so that,
\begin{equation}
-\frac{1}{D}\nabla_{\nu}M=-\frac{1}{D}\nabla_{\mu}T,
\end{equation}
which means that $M-T$ is a constant\footnote{%
When we set $f(R)=R$ and $D=4$, we find precisely the standard integrability
result of ref. \cite{weinberg89}, (note that from (\ref{Dzerotrace1}) it
follows that $M=-R$ in that case while, in general, $M=((2-D)/2)R$).}:
\begin{equation}
M-T=D\lambda.  \label{lambda}
\end{equation}
Using (\ref{lambda}), the traceless $f(R)$ equations (\ref{Dzerotrace2}),
become
\begin{equation}
{M}_{\mu \nu }-\frac{1}{D}g_{\mu \nu }M+\frac{1}{D}g_{\mu \nu }T=T_{\mu \nu
},
\end{equation}%
or finally,
\begin{equation}
{M}_{\mu \nu }-\lambda g_{\mu \nu }=T_{\mu \nu }.  \label{ltheory}
\end{equation}%
We see that traceless $f(R)$ theory, like its non-traceless version \cite%
{ba-co88}, reduces to traceless general relativity when we take $f(R)=R$.
This means that here, like in traceless general relativity \cite{einstein},
the quantity $\lambda$ is an arbitrary integration constant, and not a
fundamental parameter of the whole theory. Similarly, for the vacuum
equations (\ref{Dzerotrace2VAC}), the equivalent system (\ref{ltheory}), (%
\ref{lambda}) becomes
\begin{equation}  \label{lambdafofr}
{M}_{\mu \nu }-\lambda g_{\mu \nu }=0,\quad\text{with}\quad \lambda= M /D.
\end{equation}

\subsection{The no-scale conformal potential}

We now wish to take the vacuum, no-scale $f(R)$ theory given by the field
equation (\ref{Dzerotrace2VAC}) into the conformal frame. Before we do so,
we note a rather subtle point. Since the theory is completely equivalent to
the system (\ref{lambdafofr}), one might be tempted to think that this can
be done by simply finding the conformal picture of the first equation in (%
\ref{lambdafofr}). This equation is exactly the original $f(R)$ equation
only having the extra $\lambda $ term on the left-hand side, and so a
conformal transformation leads to the same potential as was found in Ref.
\cite{ba-co88}, but now with an extra term proportional to $\lambda $. For
example, when $D=4$ and $f(R)$ is quadratic, the potential is $V(\phi
)=e^{-\phi }\left[ (e^{2\phi }-e^{\phi })^{2}+\lambda \right]$.\footnote{%
We further note here that when calculating the slow-roll parameters in this
approach, the most dominant terms will contain a $\lambda $ term needs to be
finely tuned. For instance, for the slow-roll parameter $\epsilon $ $=\frac{1%
}{2}\left( \frac{V^{\prime }}{V}\right) ^{2}$, this dominant term reads
\begin{equation}
\frac{-2e^{-5\phi }}{1+e^{-2\phi }-2e^{-\phi }+\lambda e^{-4\phi }},
\label{dom}
\end{equation}%
and so we need to impose a fine-tuning condition on the integration constant
$\lambda $,
\begin{equation}
\lambda <e^{4\phi },
\end{equation}%
at the end of inflation, so as to ensure that the denominator in Eq. (\ref%
{dom}) tends to 1. Our previous results, in Ref. \cite{ba-co88}, about the
possibility of inflation in all spacetimes with quadratic lagrangians when $%
D=4$, and their subsequent agreement with the microwave data correspond to
the special choice of initial conditions $\lambda =0$, in the present case.}
However, this is not really correct because one has to also find the
conformal image of the integrability constraint $M=D \lambda$. This will
generate new terms which will amend the original potential, and therefore
for the current traceless case will end up with a different function than
the old one.

Instead, we have found that the simplest way to proceed is to conformally
transform directly the original equation (\ref{Dzerotrace2VAC}), and not
work with the equivalent system (\ref{lambdafofr}). To do this, we introduce
a new scalar field $\phi $ as in Eq. (\ref{def}) and following the steps
detailed in the Appendix, Part B, we find that the conformally transformed
field equation (\ref{Dzerotrace2VAC}) becomes,
\begin{equation}
\tilde{G}_{\mu \nu }=\tilde{T}_{\phi ,\mu \nu },  \label{form1}
\end{equation}%
where
\begin{equation}
\tilde{T}_{\phi ,\mu \nu }=\frac{1}{4}(D-1)(D-2)\nabla _{\mu }\phi \nabla
_{\nu }\phi -\frac{1}{2}\tilde{g}_{\mu \nu }\left( \frac{(D-1)(D-2)}{4}%
(\nabla \phi )^{2}+2V(\phi )\right) ,  \label{phi1}
\end{equation}%
with the effective scalar field potential given by
\begin{equation}
V(\phi )=\left( \frac{D-2}{4D}\right) e^{-\phi }R.  \label{V}
\end{equation}%
This is the self-interacting scalar field potential for the traceless vacuum
$f(R)$ field equations (\ref{Dzerotrace2VAC}) in the conformal frame. It has
a number of properties. Firstly, it is important to realize that this
potential is different from the standard $f(R)$ potential $%
(1/2)(Rf^{\prime}-f)f^{\prime D/(2-D)}$ found in \cite{ba-co88}. Secondly,
its sign is determined by the inversion of $f(R)$ and by using (\ref{def}),
and this in fact provides, when possible, the simplest way to obtain the
scalar curvature $R$ in the present context. Thirdly, in general, $R$
satisfies the constraint (\ref{lambda}) with $M$ given by Eq. (\ref{trace}),
which for a given function $f(R)$ gives a second order differential equation
for $R$ (in distinction to the GR case where $R$ is a constant leading to
the exponential potential (\ref{potnew})). Lastly, in the variables $%
(g,f^{\prime })$ the potential (\ref{V}) is scale-invariant (something which
is not true of the potential in the conformal transformation of the
`standard' (non-traceless) higher-order gravity theory derived in \cite%
{ba-co88}), in the sense of dimensional homogeneity, precisely like in
Section II. Therefore, like in the traceless GR case, we are allowed here to
make arbitrary linear transformations $\phi\rightarrow a\phi +b$, and shifts
in the origin of the potential, without affecting the property of
dimensional homogeneity of the field equations (\ref{form1}), (\ref{phi1}).

\subsection{$R^n$-inflation and the cosmological constant}

As an direct application of these properties, the no-scale $f(R)$-theory in
the conformal frame shares a number of novel features, different from those
of the standard $f(R)$ theory. First, it is important to notice that in all
traceless higher-order polynomial gravity theories, flat potentials occur
quite generally in the conformal traceless sub-theory. For the generic
leading-order lagrangian,
\begin{equation}
f(R)=R+AR^{n},\quad n\geq 2,  \label{dn}
\end{equation}%
solving for $R$ and using (\ref{def}) we find the potential to be:
\begin{equation}
V_n(\phi )=ce^{-\phi }\left( e^{\frac{D-2}{2}\phi }-1\right) ^{\frac{1}{n-1}%
},  \label{Vn}
\end{equation}%
where the constant $c\equiv ((D-2)/4D)(1/nA)^{1/(n-1)}$. This potential is
different from that of the standard $R^n$-inflation for any $n$ (for the
special case $D=4$, compare, eg., Eq. 10 in Ref. \cite{moto}). Taking into
account the above mentioned property of scale invariance for the conformal
traceless potential, this form leads to a huge variety of flat potentials in
all dimensions and any leading degree of the gravitational lagrangian.  We
recall that in the standard conformal correspondence between $f(R)$ gravity
theory and general relativity plus a scalar field first studied in \cite%
{ba-co88} there is a different structure: flat plateaux in $V(\phi )$ arise
in $D$ spacetime dimensions only when the highest polynomial term in $R$ in
the $f(R)$ lagrangian is $\frac{1}{2}D$, and so for $D=4$ this allows at
most a quadratic term.

By contrast, here we see that a flat potential conducive to exponential
inflation is easy to achieve for the theory in (\ref{dn}) in all dimensions:
because the potential (\ref{V}), so also (\ref{Vn}) is scale invariant, we
have the freedom to shift and stretch it (and its graph) anyway we like
without altering the theory. For example, we can move the graph to the
origin by shifting it suitably, and also replace $\phi$ by $a\phi +b$ for
any constants $a,b$. If we shift and stretch in a suitable neighborhood of
the origin, then we end up with the characteristic flat plateau for any $n$.

For example, when $D=4, n=2$, instead of the standard quadratic potential
proportional to the combination $(1-e^{-\phi})^2$, we find
\begin{equation}
V_2=\frac{1}{16A}(1- e^{-\phi}),  \label{V2}
\end{equation}
which using the scale freedom this can be written as,
\begin{equation}
V_2=a(1-g e^{-b\phi})+d ,  \label{V2a}
\end{equation}
where $a, b, d, g$ are arbitrary constants. This potential satisfies, $V(0)=d
$ and $\lim_{\phi\rightarrow\infty}V=a+d.$ It is reminiscent of the
supergravity potentials studied in Refs. \cite{eno}, Eqns. (70), (72), and
\cite{costa}, Eqn. (2.10). But here there is a crucial difference that the
potential is not an approximate form obtained after some truncation of an
asymptotic expansion like in the supergravity models, but an exact result.

Therefore, we conclude that slow-roll inflation is possible in traceless
nonlinear $f(R)$ theories. Quite a large family of underlying models lead to
observationally compatible outcomes for inflation without requiring any
fine-tuning.

A second implication is that the cosmological constant, as well as the
height of the inflationary plateau and the potential minimum (vacuum energy)
are all unrelated, arbitrary constants. In this scenario, the well-known
graceful exit problem for inflation is not well-defined because what drives
inflation is generally unrelated to the phase after it. For instance, in the
traceless $R^2$-inflation example with $D=4$, the cosmological constant is
an arbitrary constant because of the traceless constraint (\ref{lambda}),
the vacuum has energy $V(0)=d$ - another arbitrary constant, and the
inflationary plateau is at another arbitrary value, $a+d$.

This means that the cosmological constant $\lambda$ may take random values
in any causal domain in the multiverse landscape (and the same is true for
the other two parameters of the theory, $V(0)$ and the asymptotic potential
value $V_\infty=a+d$). It is like an arbitrary constant in the solution set
of a differential equation, its possible values form the set of all
solutions of the equation, the general solution. Likewise, the fact that the
cosmological constant appears in the fundamental law of the theory as an
arbitrary constant satisfying an integrability condition, leads to the
interpretation that its possible values form the set of all local domains-a
multiverse of local regions each satisfying a different law (see below) and
all of them coexisting in the `parent' set.

Of course, the fact that the cosmological constant is an arbitrary constant
in this model does not imply that it cannot be equated to its presently
observable value. This then provides an unconventional explanation of the
presently observed value of the cosmological constant: In the context of a
no-scale theory that leads to inflation, it is not a fundamental property of
the fundamental laws but a random outcome of them.

Using these results, we imagine a large number $N$ of causal domains
(`universes') such that $n_{1}$ of them have vacuum energy $\lambda _{1}$, $%
n_{2}$ have vacuum energy $\lambda _{2}$, $\dots $, etc, with the total
number of universes forming the partition
\begin{equation}
N=n_{1}+n_{2}+\cdots ,
\end{equation}%
and the total vacuum energy of the partition system (`the multiverse') is
\begin{equation}
\Lambda =n_{1}\lambda _{1}+n_{2}\lambda _{2}+\cdots .
\end{equation}%
We may import the required elements from statistical mechanics in this
scheme, with the difference that here we speak of \emph{vacuum} energies of
the domains in the partition. We assume that there is a most probable
partition in the multiverse, and when this is achieved, we say that the
multiverse is in statistical equilibrium. If we assume that this partition
is given by the Maxwell-Boltzmann distribution law, then the partition
function will be given by the form,
\begin{equation}
Z=\sum_{i}g_{i}e^{-\lambda _{i}/kT}.  \label{z}
\end{equation}%
Here there is no restriction on the number of universes that can occupy a
given vacuum energy state, there are different intrinsic probabilities $g_{i}
$ for a given vacuum state $\lambda _{i}$. From these definitions we see
that there is a temperature $T$ associated with the multiverse that is
related to the average vacuum energy of each universe ($\Lambda /N$).

Assuming that $Z$ is the Maxwell-Boltzmann distribution, we find that the
occupation numbers decrease as their vacuum energy increases. So at very low
temperatures, we expect from Eq. (\ref{z}) that only the lowest vacuum
energy states can be occupied, and the very low vacuum energies of
individual universes become more probable observationally. At higher
temperatures, higher vacuum energies become more probable, and we expect the
multiverse to become more disordered at higher temperatures (at zero
temperature only the lowest vacuum energy is occupied).

\section{Slow-roll inflation predictions for the CMB}

We can calculate the simple first-order predictions that our theory makes
for the CMB in the $(n_{s}-r)$-plane (`tilt' and `tensor-to-scalar' ratio).
We define the potential slow-roll parameters in the usual way, \cite%
{LLyth,LPB}, by
\begin{equation}
\epsilon=\frac{1}{2}\left( \frac{V^{\prime}}{V}\right) ^{2},\quad\eta =\frac{%
V^{\prime\prime}}{V},\;n_{s}=2\eta-6\epsilon  \label{SRs}
\end{equation}
and the e-folding function that gauges the length of inflation by,
\begin{equation}
N=\int_{\phi_{f}}^{\phi_{i}}\frac{1}{\sqrt{2\epsilon}}d\phi.  \label{N}
\end{equation}

\subsection{The no-scale $R+AR^2$ theory in four dimensions}

For the no-scale potential (\ref{Vn}) with $D=4,n=2$, namely, Eq. (\ref{V2}%
),
\begin{equation*}
V=\frac{1}{16A}(1-e^{-\phi}), \
\end{equation*}
when $\phi_{i}\gg\phi_{f}$, we find that
\begin{equation}
\epsilon=\frac{1}{2}e^{-2\phi},\quad\eta=-e^{-\phi},\quad N=e^{\phi },
\label{SR2}
\end{equation}
so
\begin{equation}
\epsilon=\frac{1}{2N^{2}},\quad\eta=-\frac{1}{N},
\end{equation}
and the spectral index and tensor to scalar ratio are,
\begin{equation}
1-n_{s}=\frac{2}{N},\quad r=16\epsilon=\frac{8}{N^{2}}.
\end{equation}
For instance, when $N=50$, $n_{s}=0.96$ and $r=3.2\times10^{-3}$. However,
to compare these values with the standard one, we have to normalize the
kinetic term of the scalar field. Redefining $\phi$ in Eq. (\ref{phi1}) by,
\begin{equation}  \label{red phi}
\widehat\phi=\frac{\sqrt{(D-1)(D-2)}}{2}\phi,
\end{equation}
we get the right kinetic term, and then using the second definition in Eq. (%
\ref{def}), we find,
\begin{equation}
\widehat\phi=\sqrt{\frac{D-1}{D-2}}\ln f^{\prime },
\end{equation}
which when $D=4$ gives the standard factor $\widehat\phi=\sqrt{2/3}\,\phi.$
In general, using the scale freedom the potential is given by Eq. (\ref{V2a}%
), for which the slow-roll parameters read:
\begin{eqnarray*}
\epsilon&=&\frac{1}{2}\left(\frac{Aa\lambda b}{Aa+d}\right)^2e^{-2b\phi}, \\
\eta&=&-\frac{Aa\lambda b^2}{Aa+d}\,e^{-b\phi}, \\
N&=&\frac{Aa+d}{Aa\lambda}\frac{1}{b^2}\,e^{b\phi}.
\end{eqnarray*}
Therefore we find the forms
\begin{equation}
\epsilon=\frac{1}{2b^2N^{2}},\quad\eta=-\frac{1}{b^2N},
\end{equation}
so that for the unnormalized value of $\phi$,
\begin{equation}  \label{unnorm}
r=16\epsilon=\frac{8}{b^2N^{2}},
\end{equation}
whereas for the normalized scalar field, we find that the final prediction
of the traceless $R+AR^2$ theory in four dimensions, ie, the conformal
potential (\ref{V2a}), is:
\begin{equation}  \label{norm}
r=\frac{12}{b^2N^{2}}.
\end{equation}
Notice that the arbitrary constant $b$ is the one appearing in Eq. (\ref{V2a}%
). When we put $b=1$ in Eq. (\ref{norm}) (or equivalently, when $b=\sqrt{2/3}
$ in Eq. (\ref{unnorm})) we obtain exactly the standard \cite{planck} form $%
r=12/N^2$.

This means that for this particular value $b=1$ for the normalized field,
the two predictions of the traceless and non-traceless versions of the $%
R+AR^2$ theory in four dimensions coincide. Therefore, any future prediction
of the form
\begin{equation}
r=\frac{12}{b^2N^{2}},\quad b\neq 1,
\end{equation}
will be a sign for the traceless $R+AR^2$ theory. Same conclusions hold if
we decide to use the unnormalized prediction (\ref{unnorm}), namely, $r=%
\frac{8}{b^2N^{2}},\quad b\neq \sqrt{2/3}.$ For example, $b=1/8$, gives the
main value $r=0.20$ of the BICEP2 result \cite{bicep2}, whereas the standard
\cite{planck} $R+AR^2$ theory prediction, that is when $b=\sqrt{2/3})$,
would be $r=0.005$, too small for the BICEP2 measurement.

\subsection{The no-scale $R+AR^n$ theory, general dimensions}

More generally, if we analyse Eq. (\ref{Vn}) for general $n$ and $D$ at
large $\phi$, we find that the general results which fall into two classes
according as (a) $D=2n,$ or, (b) $D\neq2n$:

(a) $D=2n:$ For $D>2$ and $n>1,$ we have%
\begin{equation}
\frac{V^{\prime }}{V}=\frac{1}{e^{(n-1)\phi }-1},
\end{equation}%
hence, we find,
\begin{eqnarray}
\epsilon  &=&\frac{1}{2[e^{(n-1)\phi }-1]^{2}}\rightarrow \frac{1}{2}%
e^{-2(n-1)\phi }, \\
\eta  &=&\frac{V^{\prime \prime }}{V}=\frac{1-(n-1)e^{(n-1)\phi }}{%
[e^{(n-1)\phi }-1]^{2}}\rightarrow -(n-1)e^{-(n-1)\phi }, \\
N &=&\int [e^{(n-1)\phi }-1]d\phi \rightarrow \frac{e^{(n-1)\phi }}{n-1},
\end{eqnarray}%
and so, to first order in slow roll,%
\begin{equation}
\epsilon \rightarrow \frac{1}{2(n-1)^{2}N^{2}},\quad \eta \rightarrow -\frac{%
1}{N},
\end{equation}%
so that
\begin{equation}
n_{s}-1=2\eta -6\epsilon =-\frac{2}{N}-\frac{3}{(n-1)^{2}N^{2}}\rightarrow -%
\frac{2}{N}.
\end{equation}%
(b) $D\neq 2n:$ In this case for large $\phi ,$ we have an asymptotically
exponential potential of the form,%
\begin{equation}
V\rightarrow c\exp \left[ \frac{(D-2n)\phi }{2(n-1)}\right] .
\end{equation}%
Hence, in that limit,%
\begin{equation}
\epsilon \rightarrow \frac{(D-2n)^{2}}{8(n-1)^{2}},\quad \eta \rightarrow
\frac{(D-2n)^{2}}{4(n-1)^{2}},
\end{equation}%
so that
\begin{equation}
n_{s}-1\rightarrow -\frac{(D-2n)^{2}}{4(n-1)^{2}}.
\end{equation}%
Introducing a new parameter, $p$, defined by
\begin{equation}
p=\frac{4(n-1)^{2}}{(D-2n)^{2}},
\end{equation}%
we have,
\begin{equation}
\epsilon =\frac{\eta }{2}=\frac{1}{2p}.
\end{equation}%
We conclude that when $D-2n\rightarrow 0$, we have $p\rightarrow \infty $,
corresponding to de Sitter spacetime asymptotically. On the other hand, when
$D-2n\sim 0$, then $p\gg 1$, and the slow-roll conditions for $\epsilon
,\eta $ are satisfied. Finally, there is the case when $D$ and $n$ take
values such that the difference $D-2n$ is far from zero. This means that the
difference $D-2n$ is neither exactly zero, nor near zero, nor asymptotic to
zero. In this case, we can always take advantage of the scaling freedom in
the potential to rescale $\phi $ and $V(\phi )$ to obtain a potential which
has a flat plateau in some finite neighborhood of zero.

We therefore conclude that in all cases the present solution to the
cosmological constant problem is confirmed because, with our results, is not
necessary anymore to tune the absolute scale of $V$ to have any special
value to achieve the `right' final state (cf. Ref. \cite{BT}, p. 438).

\section{Conclusions}

In this paper we have generalised an idea, originally due to Einstein, to
consider the traceless version of the gravitational field equations.
Einstein applied this idea to general relativity but we have extended it to $%
f(R)$ gravity theories.

In the case of traceless general relativity, we have shown that the solution
space of the vacuum theory has the property that the Weyl tensor evaluated
on solutions of the theory is covariantly constant iff the cosmological
constant is zero. This is unlike the vacuum Einstein equations where the
Weyl tensor is always covariantly constant.

In this paper we are particularly interested in those solutions of the
vacuum traceless equations having a non-constant Weyl tensor. We have shown
that under a conformal transformation the theory becomes general relativity
with a scalar field source having an exponential potential. The conformally
related version of the traceless theory has the further interesting property
that the field equations in the conformal frame are dimensionally
homogeneous. This means that we have an extra freedom to scale the conformal
potential, thus obtaining interesting inflationary solutions. Unfortunately,
the exponential potential of the conformal version of the theory has certain
undesirable features such as a graceful exit problem from inflation and also
sets incompatible constraints with respect to current CMB observations.

We then applied the previous ideas to $f(R)$ gravity and this has led to a
new version of the theory, the no-scale $f(R)$ gravity theory. This theory
has certain novel features which become apparent if one takes the traceless
version of the vacuum $f(R)$ gravity to the conformal frame. After deriving
the traceless $f(R)$ field equations for the first time in this paper, we
have shown that in vacuum these higher-order equations are conformally
equivalent to general relativity plus a self-interacting scalar field, Eqs. (%
\ref{form1})-(\ref{phi1}), with a scale-invariant potential given by Eq. (%
\ref{V}). This potential is quite distinct from the one found in the case of
standard $f(R)$ gravity, and in addition, like in the `no-scale' general
relativity case it has a scaling freedom that leads to a whole new family of
exact forms conducive to new types of cosmological evolution.

Indeed, the consideration of inflation in the framework of the no-scale $f(R)
$ gravity has a number of novel features. Perhaps the most distinctive one
is the fact that (as it follows from the analysis of the $R^n$-inflation)
the cosmological constant, the vacuum energy (the zero of the scalar field
potential), as well as the height of the inflationary plateau are all
unrelated arbitrary constants. This means that they can take on values in a
random way, and this fact leads to a new statistical interpretation of the
cosmological constant and the multiverse scenario. We believe that similar
features like those leading to a statistical interpretation of the
multiverse scenario discussed here may also arise in other versions of
modified gravity which contain the no-scale aspect considered here.

The conformal potential of the no-scale $f(R)$ gravity also gives rise to
flat plateaux which offer the possibility of exponentially rapid, slow-roll
inflation with values of $n_{s}$ and $r$ that agree well with observational
bounds from Planck \cite{planck, pico}. We have delineated the outcomes of
slow-roll inflation in the two distinctive situations defined by the
relation between the spacetime dimension and the highest polynomial order of
$f(R)$. Specialising to four-dimensional spacetimes, we have derived the
slow roll parameters for these types of inflation and shown how the results
are compatible with microwave limits on inflationary parameters without
requiring any fine-tuning or choosing special values of initial conditions.

This considerably enlarges the range of observationally acceptable
inflationary models significantly beyond those mapped in ref. \cite{pico}.
In particular, in the case of a quadratic lagrangian theory, we are led to a
generic prediction that the tensor-to-scalar ratio given by Eq. (\ref{norm})
contains the free adjustable parameter $b$ with the value $b=1$
corresponding to the standard, non-traceless quadratic theory. If future
observations prove that $b$ takes exactly the value one, then this will mean
that no-scale theories are strongly disfavored. Any other value of $b$ will
be a strong signal in favor of the novel no-scale aspects of gravity
discussed in this paper.

At the end of this inflationary phase, the separation of the cosmological
constant term from any quantum vacuum energy density means that it can be
arbitrarily small, and remain small for billions of years, consistent with
current observations of our local domain.

\begin{acknowledgments}
\noindent JDB is supported by the Science and Technology Funding Council (STFC) of the UK. SC would like to thank the University of the Aegean, Greece,  CERN, Switzerland, NTUA, Greece and AUM, Kuwait for support during the years this work was done.
\end{acknowledgments}

\appendix

\section{The conformal, scale-invariant potential}

In this Appendix, we provide a full derivation of the conformal,
scale-invariant potentials (\ref{potnew}), (\ref{V}). In the Part A1, we
consider the derivation in the context of the traceless Einstein equations;
while in Part A2 we derive the potential for the traceless $f(R)$ theory.

For the conformal transformation (\ref{def}), we use the standard relations
for the Ricci tensor,
\begin{align}
\tilde{R}_{\mu\nu} & =R_{\mu\nu}+\frac{D-1}{D-2}(f^{\prime})^{-2}\nabla
_{\mu}f^{\prime}\nabla_{\nu}f^{\prime}  \notag \\
& -(f^{\prime})^{-1}\nabla_{\mu}\nabla_{\nu}f^{\prime}-\frac{1}{D-2}g_{\mu
\nu}(f^{\prime})^{-1}\Box f^{\prime},  \label{ricci}
\end{align}
and the scalar curvature,
\begin{equation}
\tilde{R}=(f^{\prime})^{\frac{2}{2-D}}\left( R+\frac{D-1}{D-2}(f^{\prime
})^{-2}(\nabla f^{\prime})^{2}-\frac{2(D-1)}{D-2}(f^{\prime})^{-1}\Box
f^{\prime}\right) ,  \label{tilde r}
\end{equation}
where $(\nabla f^{\prime})^{2}\equiv g^{\rho\sigma}\nabla_{\rho}f^{\prime
}\nabla_{\sigma}f^{\prime}$.

\subsection{No-scale Einstein theory}

Using Eq. (\ref{tr0ein}) and the transformations (\ref{ricci}), (\ref{tilde
r}), we can write the relation of the conformally related traceless Einstein
tensors,
\begin{align}
\widetilde{\widehat{G}}_{\mu\nu} & =\widehat{G}_{\mu\nu}+\frac{D-1}{D-2}%
(f^{\prime})^{-2}\nabla_{\mu}f^{\prime}\nabla_{\nu}f^{\prime}-\frac {D-1}{%
D(D-2)}g_{\mu\nu}(f^{\prime})^{-2}(\nabla f^{\prime})^{2}  \notag
\label{conf tr1 ein} \\
& -(f^{\prime})^{-1}\nabla_{\mu}\nabla_{\nu}f^{\prime}+\frac{1}{D}g_{\mu\nu
}(f^{\prime})^{-1}\Box f^{\prime}.
\end{align}
(Tracing both sides we recover `$0=0$' as expected.) Since the original
theory $\widehat{G}_{\mu\nu}=0$ is scale invariant, we can use the \emph{%
harmonicity condition} (cf. \cite{wald}, Eq. D. 13) for $f^{\prime}$,
\begin{equation}
\Box f^{\prime}-\frac{D-2}{4(D-1)}Rf^{\prime}=0,
\end{equation}
to rewrite the last term in (\ref{conf tr1 ein}), so that using the field
equation $\widehat{G}_{\mu\nu}=0$, Eq. (\ref{conf tr1 ein}) becomes,
\begin{align}
\widetilde{\widehat{G}}_{\mu\nu} & =\frac{D-1}{D-2}(f^{\prime})^{-2}\nabla_{%
\mu}f^{\prime}\nabla_{\nu}f^{\prime}-\frac{D-1}{D(D-2)}g_{\mu\nu
}(f^{\prime})^{-2}(\nabla f^{\prime})^{2}  \notag  \label{conf tr2 ein} \\
& -(f^{\prime})^{-1}\nabla_{\mu}\nabla_{\nu}f^{\prime}+\frac{D-2}{4D(D-1)}%
g_{\mu\nu}R.
\end{align}
In terms of the scalar field $\phi,$ we find that
\begin{equation}
\widetilde{\widehat{G}}_{\mu\nu}+\frac{D-2}{2}\nabla_{\mu}\nabla_{\nu}\phi=%
\tilde{T}_{\phi,\mu\nu},  \label{vac conf TEE}
\end{equation}
where
\begin{equation}
\tilde{T}_{\phi,\mu\nu}=\frac{D-2}{4}\left[ \nabla_{\mu}\phi\nabla_{\nu}\phi-%
\frac{D-1}{D}\tilde{g}_{\mu\nu}(\nabla\phi)^{2}+\frac{1}{D(D-1)}\tilde {g}%
_{\mu\nu}e^{-\phi}R,\right]  \label{matter tensor 1}
\end{equation}
which is the field equation (\ref{vac conf TEE1}), (\ref{matter tensor 2}).
Here, we have used the following results which follow from the definition of
the conformally related metrics: from (\ref{def}), $f^{\prime}=e^{\frac{D-2}{%
2} \phi}$, and so the logarithmic derivative of $f^{\prime}$ is $\nabla_{\mu
}f^{\prime}/f^{\prime}=(D-2)/2\,\nabla_{\mu}\phi$. This gives
\begin{equation}
\nabla_{\mu}\left( \frac{\nabla_{\nu}f^{\prime}}{f^{\prime}}\right) =\frac{%
D-2}{2}\nabla_{\mu}\nabla_{\nu}\phi.
\end{equation}
Calculating again the covariant derivative of the log derivative of $%
f^{\prime}$ using the quotient rule, we find
\begin{equation}
f^{\prime-1}\nabla_{\mu}\nabla_{\nu}f^{\prime}=\frac{D-2}{2}%
\nabla_{\mu}\nabla_{\nu}\phi+\left( \frac{D-2}{2}\right)
^{2}\nabla_{\mu}\phi\nabla _{\nu}\phi.
\end{equation}
Substituting in Eq. (\ref{conf tr2 ein}) and expressing the first two terms
on the right-hand side in terms of $\phi$, we arrive at Eq. (\ref{vac conf
TEE}).

\subsection{No-scale $f(R)$ theory}

First, write the vacuum equation (\ref{Dzerotrace2VAC}) in a form containing
the Einstein tensor. Since
\begin{equation}
\widehat{G}_{\mu\nu}=R_{\mu\nu}-\frac{1}{D}g_{\mu\nu}R=G_{\mu\nu}-\frac {2-D%
}{2D}g_{\mu\nu}R,
\end{equation}
and assuming that $f^{\prime}$ has constant sign, substituting in equation (%
\ref{Dzerotrace1}) we find
\begin{equation}
G_{\mu\nu}-(f^{\prime})^{-1}\nabla_{\mu}\nabla_{\nu}f^{\prime}+\frac{1}{D}%
g_{\mu\nu}(f^{\prime})^{-1}\left( \Box f^{\prime}-\frac{2-D}{2}f^{\prime
}R\right) =0.  \label{np1}
\end{equation}
We then take this equation to the conformal frame. We do this in two steps,
first expressing everything in terms of $f^{\prime}$. The conformal relation
of the tilded Einstein tensor to the untilded one is found using the
transformations (\ref{ricci}), (\ref{tilde r}),
\begin{align}
\tilde{G}_{\mu\nu} & =G_{\mu\nu}+\frac{D-1}{D-2}(f^{\prime})^{-2}\nabla
_{\mu}f^{\prime}\nabla_{\nu}f^{\prime}-\frac{D-1}{2(D-2)}g_{\mu\nu}(f^{%
\prime })^{-2}(\nabla f^{\prime})^{2}  \notag  \label{conf tr ein} \\
& -(f^{\prime})^{-1}\nabla_{\mu}\nabla_{\nu}f^{\prime}+g_{\mu\nu}(f^{\prime
})^{-1}\Box f^{\prime}.
\end{align}
Solving this for the untilded Einstein tensor $G_{\mu\nu}$ and substituting
back to the field equation (\ref{np1}), we obtain
\begin{align}
\tilde{G}_{\mu\nu} & -\frac{D-1}{D-2}(f^{\prime})^{-2}\nabla_{\mu}f^{\prime
}\nabla_{\nu}f^{\prime}+\frac{D-1}{2(D-2)}g_{\mu\nu}(f^{\prime})^{-2}(\nabla
f^{\prime})^{2}  \notag \\
& +\frac{1-D}{D}g_{\mu\nu}(f^{\prime-1}\Box f^{\prime}-\frac{2-D}{2D}%
g_{\mu\nu}R=0.  \label{conf trace f'}
\end{align}
Using the scalar field $\phi$ from Eq. (\ref{def}) in the field equation (%
\ref{conf trace f'}), we can express the second and third terms in the form
\begin{equation}
-\frac{(D-1)(D-2)}{4}\nabla_{\mu}\phi\nabla_{\nu}\phi+\frac{(D-1)(D-2)}{8}%
g_{\mu\nu}({\nabla\phi})^{2}.  \label{phi terms}
\end{equation}
We can also simplify the last two terms,
\begin{equation}
g_{\mu\nu}\left( \frac{1-D}{D}(f^{\prime-1}\Box f^{\prime}-\frac{2-D}{2D}%
R)\right) ,
\end{equation}
by noticing that since the original theory (\ref{Dzerotrace2}) is scale
invariant, so we can use the \emph{harmonicity condition} (cf. \cite{wald},
Eq. D. 13) for $\phi,$ (or $f^{\prime}$),
\begin{equation}
\Box f^{\prime}-\frac{D-2}{4(D-1)}Rf^{\prime}=0,
\end{equation}
to write these two terms in the form
\begin{equation}
\frac{D-2}{4D}g_{\mu\nu}R.  \label{potential term}
\end{equation}
Substituting Eqns. (\ref{phi terms}), (\ref{potential term}) into the field
equation (\ref{conf trace f'}), we find
\begin{equation}
\tilde{G}_{\mu\nu}=\frac{(D-1)(D-2)}{4}\nabla_{\mu}\phi\nabla_{\nu}\phi -%
\frac{(D-1)(D-2)}{8}g_{\mu\nu}({\nabla\phi})^{2}-\frac{D-2}{4D}g_{\mu\nu}R,
\end{equation}
which gives (\ref{form1})-(\ref{V}), after we express the second term on the
right-hand side using the identity $g_{\mu\nu}({\nabla\phi})^{2}=\tilde {g}%
_{\mu\nu}({\tilde{\nabla}\phi})^{2}$ (which holds when multiplying and
dividing by $e^{\phi}$ and noting that $e^{-\phi}(\nabla\phi)^{2}=e^{-\phi
}g^{\rho\sigma}\nabla_{\rho}\phi\nabla_{\sigma}\phi=\tilde{g}^{\rho\sigma
}\nabla_{\rho}\phi\nabla_{\sigma}\phi$), and likewise for the last term, $%
g_{\mu\nu}R=\tilde{g}_{\mu\nu}e^{-\phi}R.$

\end{document}